\documentclass[12pt]{iopart}

\usepackage{graphicx}
\usepackage{color}
\usepackage{epstopdf}
\usepackage{listings}
\usepackage{braket}
\usepackage{hyperref}

\usepackage{bm}
\usepackage{cite}
\usepackage{iopams}
\usepackage[detect-all]{siunitx}

\newcommand{\Range}[2]{\numrange[range-phrase = \text{--}]{#1}{#2}}
\newcommand{\reffig}[1]{Fig.~\ref{#1}}
\newcommand{\refeq}[1]{Eq.~(\ref{#1})}
\newcommand{\refeqs}[2]{Eqs.~(\ref{#1})-(\ref{#2})}

\begin{document}

\title{Excitation of knotted vortex lines in matter waves}

\author{F. Maucher$^{1,2}$, S. A. Gardiner$^1$, I. G. Hughes$^1$}

\address{$^1$Joint Quantum Centre (JQC) Durham-Newcastle, Department of Physics, Durham University, Durham DH1 3LE, United Kingdom\\
$^2$ Department of Mathematical Sciences, Durham University, Durham DH1 3LE, United Kingdom
}

\ead{fabian.maucher@durham.ac.uk}

\begin{abstract}  
We study the creation of knotted ultracold matter waves in Bose-Einstein condensates via
coherent two-photon Raman transitions with a $\Lambda$ level configuration.  
The Raman transition allows an indirect transfer of atoms from the internal state $\ket{a}$ to the target state $\ket{b}$ via an excited state $\ket{e}$, that
would be otherwise dipole-forbidden. 
This setup enables us to imprint three-dimensional knotted vortex lines embedded in the 
probe field to the density in the target state.
We elaborate on experimental feasibility as well as on subsequent dynamics of the matter wave. 
\end{abstract}

\maketitle

\section{Introduction}

Waves can exhibit wavefront dislocations or vortices, i.e. phase 
singularities of the complex wavefuntion, where the modulus of the wave
vanishes, and around which the phase of the wave changes by a multiple of 
2$\pi$~\cite{Berry:royal:1974}. 
In~\cite{Irvine:NatPhys:2008}, it has been shown that Maxwell's equations in fact admit solutions where 
these lines of dislocation or vortex lines are tied into knots embedded in light fields.
Here, the term ``knot'' refers to an embedding of a circle $S^1$ into a three-sphere $S^3$~\cite{knot_book}.
Both theoretical~\cite{Shabtay:OC:2003} as well as 
experimental advances~\cite{Padgett:NJP:2005,Whyte:NJP:2005,Dennis:Nature:2010,Shanblatt:OE:11} in three-dimensional light shaping allowed to 
the experimental realization of knotted vortex lines embedded in optical fields. 
Apart from optics, the study of knotted topological defect lines and their dynamics has fascinated scientists from diverse settings, including
classical fluid dynamics~\cite{Moffatt:JFM:1969,Moffatt:nature:1990}, excitable media~\cite{Paul:PRE:2003,Sutcliffe:PRL:2016}, 
chiral nematic colloids~\cite{Tkalec:science:2011} to semiconductors~\cite{Babaev:PRL:2002,Babaev:PRB:2009}.

Lord Kelvin speculated in 1867~\cite{Kelvin:1867} that atoms can be described by vortex tubes in the aether.
In fact, almost two decades ago it was suggested that (nontrivial) knots might exist as stable solitons or Hopfions
in three-dimensional field theories~\cite{Faddeev:Nature:1997}. A Skyrme model served as an example for a field theory 
which admits stable knot solitons~\cite{Paul:PRL:1998}. 

In the context of spinor Bose-Einstein condensates (BECs)~\cite{TinLun:PRL:98,Machida:JphysJpn:1998}, the
existence of knots with nonzero Hopf charge was studied~\cite{Niemi:PRB:2002,Ueda:PRL:08,Ueda:PTPS:2010} and recently realized
experimentally~\cite{Mottonen:NatPhys:2016}. However, there is work~\cite{Speight:JGP:2010} that casts doubt on existence of
stable Hopfions in two-component Ginzburg-Landau type systems. 

Contrasting these studies, 
numerical investigations on the robustness and centre of mass motion of {\em knotted vortex lines} embedded in a {\em single}
component BEC have been undertaken~\cite{Barenghi:PRE:2012,Barenghi:JOP:2014}. 
Such formations are typically unstable, since there is no topological stabilization mechanism and reconnections of 
vortex lines are allowed due to the occurrence of the quantum stress tensor in the hydrodynamic formulation of the Gross-Pitaevskii Equation. 
Nucleation~\cite{Frisch:PRL:1992} and reconnection of vortex lines~\cite{Koplik:PRL:1993,Bewley:PNAS:2008,Barenghi:PhysFluid:2012} and vortex line bundles~\cite{Barenghi:PRL:2008} 
and subsequent emission of sound waves~\cite{Leadbeater:PRL:2001} have been theoretically studied extensively. 

In this paper, we follow up on the idea formulated in~\cite{Ruostekoski:PRA:2005}, and 
discuss a general experimental scheme to create a two-component BEC which contains a knotted vortex line in one 
of its components. 
We suggest using a light field containing a knotted vortex line as probe field of a Raman-pulse that drives a 
coherent two-photon Raman transition of three-level atoms with $\Lambda$-level configuration [cf.~\reffig{fig:lambda_scheme}(a)]. 
Previously, similar methods have been used to create dark solitons and vortices~\cite{Wright:PRL:1997,Dum:prl:98,Ruostekoski:PRL:2004,Ruostekoski:PRA:2005,Andersen:PRL:2006} 
and have been experimentally realized using microwave~\cite{Cornell:PRL::1999_2,Cornell:PRL::1999} and more recently optical coupling~\cite{Schmiegelow:ARXIV:2015}. 
The pump beam will be the mentioned knotted (stationary) light beam, whereas 
the control beam is a co-propagating plane wave to remove fast oscillations in the $z$-direction~\cite{Ruostekoski:PRA:2005}.
The large controllability of the pulse parameters (i.e. strength and duration of the Rabi-pulse) 
allows for a large controllability of the excitation. 
The dynamics of two-component BECs can be monitored in real time via in situ measurements without ballistic expansion~\cite{Cornell:PRL:2000}. 
Using numerical methods, we study excitation and subsequent dynamics of specific examples of knotted matter waves for experimentally feasible parameters. 

This paper thus paves the way for experimentally accessing many of the 
phenomena discussed only theoretically in, e.g.~\cite{Barenghi:PRE:2012,Barenghi:JOP:2014,Irvine:arXiv:2015},
and if such system were realized experimentally, it would give controlled experimental access to reconnection of vortex lines,  
subsequent emission of sound waves and more generally quantum turbulence. 

\section{Model}\label{sec:artificial_light}
\subsection{Equations of motion for light-matter wave coupling}

\begin{figure}\begin{center}
\includegraphics[width=\textwidth]{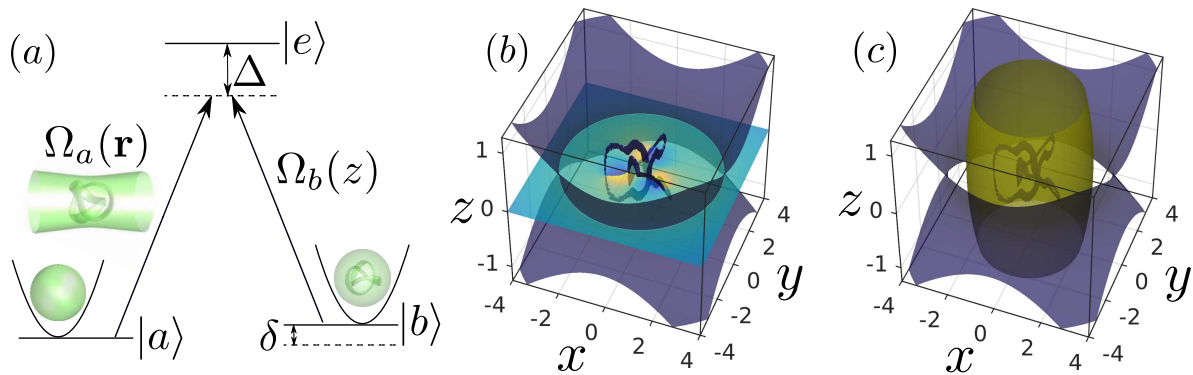}
\caption{
(color online) (a) Schematic sketch of a Raman-type transition with $\Lambda$-configuration. The state $\ket{a}$ is
off-resonantly coupled to state $\ket{b}$ with two-photon detuning $\delta$.
The knotted light field $\mathcal{E}$ proportional to $\Omega_a({\bf r})$ off-resonantly couples state $\ket{a}$ to $\ket{e}$ with detuning $\Delta$. 
The final state $\ket{b}$ then reflects the involved structure of the light field $\mathcal{E}$ associated with $\Omega_a({\bf r})$.
(b) depicts an isosurface of the intensity (purple) of the light-field $\mathcal{E}$ at a low isointensity value and a slice of its 
phase in the $(x,y)$-plane. Additional to the vortex lines in the center of the beam, the usual diffraction cones become visible.
(c) Illustration of the situation at $t=0$ for the parameters discussed in~\ref{sec:trefoil}.
The isodensity of the cigar-shaped wave function $\psi_a$ (yellow) is subject to the knotted light field (purple) from Fig. (b).
\label{fig:lambda_scheme}}
\end{center}\end{figure}

Consider a three-level atom in $\Lambda$-configuration with ground states $\ket{a}$ and $\ket{b}$, which 
are off-resonantly coupled to an excited state $\ket{e}$ with detuning $\Delta$ and spatially dependent Rabi-frequencies
$\Omega_a({\bf r})$ and $\Omega_b(z)$. Assuming $\Delta\gg\Omega_a,\Omega_b$, the excited state can be 
adiabatically eliminated, and the dynamics of the condensate wave function confined in a trap $V({\bf r})$ 
can be described~\cite{Dum:prl:98,Ruostekoski:PRA:2005} using~\refeqs{eq:eqmo1}{eq:eqmo2}
\begin{eqnarray}
\fl i\hbar\partial_t\psi_a&=\left(-\frac{\hbar^2}{2m}\nabla^2 + V({\bf r}) + Ng_{aa} |\psi_a|^2+Ng_{ab}|\psi_b|^2\right)\psi_a+\frac{\hbar\Omega_a({\bf r})\Omega_b^*(z)}{8\Delta}\psi_b\label{eq:eqmo1}\\
\fl i\hbar\partial_t\psi_b&=\left(-\frac{\hbar^2}{2m}\nabla^2 + V({\bf r}) + Ng_{ba} |\psi_a|^2+Ng_{bb}|\psi_b|^2 -\hbar\delta \right)\psi_b+\frac{\hbar\Omega_a^*({\bf r})\Omega_b(z)}{8\Delta}\psi_a
 \label{eq:eqmo2}
\end{eqnarray}
with $g_{kj}=4\pi\hbar^2 a_{kj}/m$, $a_{kj}$ denoting the scattering length between the species and 
\begin{equation}
V({\bf r})= \frac{1}{2}m\omega^2 r_\perp^2+\frac{1}{2}m\omega_z^2 z^2.
\end{equation}
\reffig{fig:lambda_scheme}(a) depicts the level scheme of three-level atoms in $\Lambda$-type configuration.
The Rabi-frequencies $\Omega_k$, $k=a,b$ relate to their corresponding electric fields via
\begin{equation}
  \Omega_k=\frac{dE_k}{\hbar},
 \label{eq:Omega_a}
\end{equation}
where the transition dipole element $d$ has been introduced. 
As electric fields, we consider a monochromatic, quasi-linearly polarized light field with 
wavevector $k_0$. 
Then, its slowly varying envelope~$\mathcal{E}$ is defined by 
\begin{equation}
 E({\bf r},t)=\sqrt{\frac{\omega_0\mu_0}{2k_0}}\mathcal{E}({\bf r})e^{i\left(k_0z-\omega_0t\right)}+{\rm c.c.},
 \label{eq:sve}
\end{equation}
where ${\rm c.c}$ denotes the complex conjugate. 
Within the paraxial approximation, the slowly varying envelope $\mathcal{E}$ of the beam is governed by 
\begin{equation}
2ik_0\partial_z \mathcal{E}({\bf r}_{\perp},z) = -\nabla^2_{\perp}\mathcal{E}({\bf r}_{\perp},z),
\label{eq:paraxial}
\end{equation}
where ${\bf r}_\perp=(x,y)$, and $\nabla^2_\perp$ denotes the transverse Laplacian.

Such a system can be realized by considering the two hyperfine ground states $\ket{a}=\ket{S_{1/2},F=2,M_F=-1}$, $\ket{b}=\ket{S_{1/2},F=2,M_F=1}$ 
of ${}^{87}$Rb, that are off-resonantly coupled to an excited state manifold. 
Then, the scattering lengths between the species $a_{kj}$ are given by $a_{ba}=5.5$nm, and the ratio $a_{aa}$:$a_{ba}$:$a_{bb}$ is given by $1.03$:$1$:$0.97$~\cite{Cornell:PRL:1998}. 
The motivation for assuming cylindrical symmetry of the trapping potential $V({\bf r})$ is associated with 
the specific form of the paraxial wave equation~\refeq{eq:paraxial}. 
Whereas we use different beams for trapping and Raman transition, the aspect ratio $\omega_z/\omega$ 
for our otherwise independent cigar shaped trap cannot be chosen arbitrarily.
Instead, in order to accommodate the optical knot in the BEC (see~\reffig{fig:lambda_scheme}(c)), 
we have to make sure that extent of the BEC in the z-direction is large enough. 
The aspect ratio of the optical beam (and the optical knot) can be expressed using the ratio between the Rayleigh 
length $z_r=k_0\sigma^2$ and the width $\sigma$ of the light field in the 
$(x,y)$-plane:
\begin{equation}
 \frac{z_r}{\sigma}=\frac{k_0\sigma^2}{\sigma}=k_0\sigma.
 \label{eq:aspect_ratio}
\end{equation}
The basic idea towards imprinting knotted vortex lines into BECs is depicted in~\reffig{fig:lambda_scheme}(a).
Consider the case where all atoms are initially in state $\ket{a}$ [$\psi_b(t=0)=0$].
Then, assuming $\Omega_i\ll \Delta$ and $\delta\approx 0$ the 
Rabi pulse coherently transfers population from $\ket{a}$ to $\ket{b}$ as 
illustrated in according to~\refeqs{eq:eqmo1}{eq:eqmo2}. The light field is depicted in~\reffig{fig:lambda_scheme}(b) and together with the density of $\psi_a$ in~\reffig{fig:lambda_scheme}(c).
The laser associated with $\Omega_b$ is chosen to be a simple plane wave, which coherently co-propagates with $\mathcal{E}$, 
so that the fast variations $\sim e^{ik_0z}$ in~\refeq{eq:sve} are canceled whenever products $\Omega_a\Omega_b^*$, as in~\refeqs{eq:eqmo1}{eq:eqmo2}, occur.
The explanation as to why such a setup allows to imprint the phase of the structured light field $\Omega_a$ onto the condensate is the following.
For short time and two-photon detuning $\delta=0$, the small change $\delta\psi_b$ to $\psi_b$ is 
given by 
\begin{equation}
 \delta\psi_b \approx-\frac{i}{\hbar}\frac{\hbar\Omega_a^*({\bf r})\Omega_b(z)}{8\Delta}\psi_a(t=0)\delta t.
 \label{eq:simple_argument}
\end{equation}
Thus, we may conclude that at least for small times, it should be possible to populate state $\ket{b}$ 
with a given phase profile using our light field. 
Once the phase profile $\theta$ has been imprinted ($\mathcal{E}=|\mathcal{E}|e^{i\theta}$) onto 
the condensate, the atoms will display motion according to ${\bf v}=\hbar\nabla\theta/m$~\cite{Ruostekoski:PRA:2005}. 
To realize such a scenario, the intensity of the involved 
laser beams must be sufficiently strong, such that the time-scales of the imprinting are small compared to
time-scales of the dynamics of the condensate.
We will use numerical methods and realistic experimental 
parameters to extend this simple idea beyond the perturbative limit and 
study its subsequent dynamics.

\subsection{Knotted Light Field}\label{sec:knotted_light_field}

Laguerre-Gaussian (LG) functions  $\mathrm{LG}_{l,p}$ 
form a basis set for solutions for the paraxial wave equation~\refeq{eq:paraxial}, and are given by the expression
\begin{eqnarray}
  \mathrm{LG}_{l,p}^{\sigma,z_r}({\bf r}_\perp,z)=&\sqrt{\frac{p!}{\pi (|l|+p)!}}\frac{r_\perp^{|l|}e^{il\varphi}}{\sigma^{|l|+1}}\frac{(1-iz/z_r)^p}{(1+iz/z_r)^{p+|l|+1}}\nonumber\\
   & \times e^{-r_{\perp}^{2}/2\sigma^2(1+iz/z_r)} L^{|l|}_p \left( \frac{r_{\perp}^{2}}{\sigma^2\left[1+\left(z/z_r\right)^2\right]} \right)
   \label{eq:LG_modes}
\end{eqnarray}
with $r_{\perp}^2=x^2+y^2$ and $L^{|l|}_p$ being the associated Laguerre polynomials. 

We seek a linear superposition of LG-modes, $\sum_{l,p}a_{lp}\mathrm{LG}_{l,p}$, 
that describe light-beams containing knotted vortex lines. 
To this end, we will review the method proposed in~\cite{Dennis:Nature:2010}, which uses Milnor polynomials~\cite{Milnor_book} as an ansatz for complex light fields
in the shape of torus knots to determine appropriate amplitudes $a_{lp}$, and rescale for application to our setup.

The basic idea~\cite{King:thesis,Dennis:Nature:2010} is to parametrize an $N$-strand braid as the roots of the polynomial 
\begin{equation}
 p_h^{N,n}(u)=\prod_{j=0}^{N-1} \left[u-s_j(h)\right]
\end{equation}
Here, $h$ denotes the height of the periodic braid and $N$ denotes the number of strands or roots of $p_h(u)$ in $u$.
Let us choose $N=2$ and~\cite{King:thesis} 
\begin{equation}
 s_j(h)=\cos(h_j)+i\sin(h_j), \quad h_j=(h-2\pi j/n)n/2.
\end{equation}
The projection of the braid onto the $(h=0)$-plane leads to a circle. The parameter $n$ represents the number of braid crossings. 
We will consider the cases $n=2,3$ in the following. 
A small computation allows us to find an explicit expression for the Milnor polynomial $p_h$
in the variables $u$ and $\exp(ih)=:v$,
\begin{equation}
 p_h^{2,n}=u^2-v^n
 \label{eq:milnor}
\end{equation}
We can now imagine a cylinder containing the braid, and ``glue'' top and bottom surfaces together to obtain a knot. In fact, one can show~\cite{Alexander:PNAS:1923} that
any knot can be represented as a closure of a braid. 
An easy way to deform our cylinder into a torus and to make it explicitly dependent on ${\bf{r}^\prime}=(x^\prime,y^\prime,z^\prime)$ 
is to write $u$ and $v$ as an inverse 
stereographic projection from three-dimensional space to a 
unit three-sphere, $\mathbb{R}^3\rightarrow\mathbb{S}^3$, i.e.
\begin{eqnarray}
u&=\frac{r^{\prime 2}-1+2iz^{\prime}}{r^{\prime 2}+1},\\
v&=\frac{2(x^{\prime}+iy^{\prime})}{r^{\prime 2}+1}\label{eq:stereo}
\end{eqnarray}
Here, $r^\prime=\sqrt{x^{\prime 2}+y^{\prime 2}+z^{\prime 2}}$ and the units $\bf{r}^\prime$ are non-dimensional. 
One can easily show that $|u|^2+|v|^2=\Re(u)^2+\Im(u)^2+\Re(v)^2+\Im(v)^2=1$, 
and thus~\refeq{eq:stereo} indeed represents a parametrisation of a three-sphere. 
Since we aim to describe an actual light field, i.e. a solution to the paraxial wave equation, 
instead of considering the Milnor polynomial~\refeq{eq:milnor} $p_h$ as it is, it is reasonable to get rid of the denominator in 
$p_h$ and to consider~\cite{Dennis:Nature:2010,King:thesis} 
\begin{eqnarray}
\xi_a({\bf r}^\prime)&=\left(u^2-v^n\right)\left(r^{\prime 2}+1\right)^n.
\label{eq:ansatz_light}
\end{eqnarray}
For $n=2,{\ }3$,~\refeq{eq:ansatz_light} 
describes a Hopf link and a trefoil knot, respectively, which 
represent the simplest non-trivial examples of a link and a knot, respectively.
The latter two will be used as exemplary fields in the following. 

In order to use~\refeq{eq:ansatz_light} to find appropriate coefficients $a_{l,p}$, 
let us rescale the light field~\refeq{eq:LG_modes} by $R_s$ to nondimensional units $(x^\prime,y^\prime,z^\prime)$, such that 
\begin{equation}
 {\bf r}_\perp={\bf r}^\prime_\perp R_s,\quad
 \frac{z}{z_r}=\frac{z/R_s}{k_0R_s\sigma^2/R_s^2}=\frac{z^\prime}{(k_0R_s) (\sigma^2/R_s^2)},
\end{equation}
to equate the light field to our knot in the same dimensionless units. 
Effectively, $R_s$ scales the abstract unit sphere to have a transverse extent of approximately $R_s$ with respect to the laser beam.
Hence, there are two length scales of our system, $R_s$ and $\sigma$, that are associated with the nodal lines of the knot and the transverse extent of the beam. 
As we will see, the ratio $w=\sigma/R_s$ will play a crucial role as an important degree of freedom, that allows us to change the width of the 
beam relative to the positions of the nodal lines of the knot. 
To find appropriate superpositions of ${\mathrm{LG}}$-modes, it is sufficient to restrict considerations to the $(x,y)$-plane only.
Let us equate~\refeq{eq:ansatz_light} with a superposition of the above-mentioned rescaled version of~\refeq{eq:LG_modes}: 
\begin{eqnarray}
\xi_a({\bf r}_\perp^\prime,z^\prime=0) = \sum_{l,p} a_{l,p}(w)\mathrm{LG}_{l,p}^{w,k_0R_sw^2}\left({\bf r}_\perp^\prime,z^\prime=0\right)\sqrt{\pi}e^{r_{\perp}^{\prime 2}/2w^2}w.
\label{eq:determine_alp}
 \end{eqnarray}
Comparing different powers in $r_\perp^\prime$ allows us to determine the finite number of coefficients $a_{l,p}$ uniquely,
which depend only on the real number $w$. Whereas this ansatz can be used for some knots, there are counterexamples, and 
there is no rigorous general proof as to why and in which cases this ansatz leads to success. 
A large value of $w$ ensures that the vortex lines 
are actually embedded in the beam, and not chopped off.
Furthermore, for finite $w$, additional vortex lines in the shape of hairpins appear for larger $z$-values, which leads to the fact that 
$w$ must be chosen large enough. On the other hand, if $w$ is too large, 
the polynomial increase in the polynomials describing the knots will not be attenuated quickly enough
by the Gaussian, and thus intensity variations become huge, which is undesirable for our setup. 
It is possible~\cite{Dennis:Nature:2010} to further optimize these coefficients to separate vortices with regions of 
larger intensity. 
Once appropriate coefficients $a_{l,p}$ have been found, the light field can be 
written down as superposition of LG modes by rescaling back
into physical units. For the sake of clarity, let us introduce an auxiliary function $f$ defined as 
$f(r_\perp/\sigma,z/z_r)/\sigma:=\mathrm{LG}_{l,p}^{\sigma,z_r}({\bf r}_\perp,z)$. Then, we find 
\begin{eqnarray}
  \mathcal{E}_a({\bf r}_\perp,z)&=\frac{A}{R_s}\sum_{l,p}a_{l,p}(w)\mathrm{LG}_{l,p}^{w,k_0R_sw^2}({\bf r}_\perp^\prime,z^\prime),\\
  &=\frac{A}{R_s}\sum_{l,p}a_{l,p}(w)\frac{1}{w}f\left(\frac{r_\perp^\prime}{w},\frac{z^\prime}{k_0R_sw^2}\right),\\
  &=A\sum_{l,p}a_{l,p}(w)\frac{1}{\sigma}f\left(\frac{r_\perp}{\sigma},\frac{z}{z_r}\right)\\
  &=A\sum_{l,p}a_{l,p}(w)\mathrm{LG}_{l,p}^{\sigma,z_r}({\bf r}_\perp,z).
  \label{eq:superpos}
\end{eqnarray}
Here, the amplitude $A$ of the light field has been introduced, which gives the intensity of the beam the right value in 
appropriate units. 
Note, that~\refeq{eq:superpos} is no longer dependent on the choice of $R_s$. 

\subsection{Rescaling}

Let us rescale~\refeqs{eq:eqmo1}{eq:eqmo2} by 
introducing spatial $r^\prime$ and time $t^\prime$ coordinates rescaled 
by oscillator length $a_0=\sqrt{\hbar/m\omega}$ and trapping frequency $\omega$, respectively,
\begin{eqnarray}
 t^\prime&=\omega t,\\
 r^\prime&=\frac{r}{a_0},\\
 \psi^\prime&=a_0^{3/2}\psi.
\end{eqnarray}
Then, after multiplying~\refeqs{eq:eqmo1}{eq:eqmo2} by $1/m\omega^2\sqrt{a_0}$, ~\refeqs{eq:eqmo1}{eq:eqmo2} become
\begin{eqnarray}
 \fl i\partial_{t^\prime}\psi_a^\prime&=\left(-\frac{1}{2}\nabla^{\prime 2} +  \frac{r_\perp^{\prime 2}}{2} + \gamma_z^2  \frac{z^{\prime 2}}{2} + \kappa_{aa} |\psi_a^\prime|^2+\kappa_{ab}|\psi_b^\prime|^2\right)\psi_a^\prime
 +\frac{\Omega_a({\bf r}^\prime)\Omega_b^*(z^\prime)}{8\omega\Delta}\psi_b^\prime\label{eq:rescaled_eqmo1}\\
 \fl i\partial_{t^\prime}\psi_b^\prime&=\frac{\Omega_a^*({\bf r}^\prime)\Omega_b(z^\prime)}{8\omega\Delta}\psi_a^\prime+ 
 \left(-\frac{1}{2}\nabla^{\prime 2} +  \frac{r_\perp^{\prime 2}}{2} + \gamma_z^2  \frac{z^{\prime 2}}{2} + \kappa_{ba} |\psi_a^\prime|^2+\kappa_{bb}|\psi_b^\prime|^2-\frac{\delta}{\omega}\right)\psi_b^\prime
 \label{eq:rescaled_eqmo2}
\end{eqnarray}
where we left away the prime for our non-dimensional time and space variables and introduced $\gamma_z=\omega_z/\omega$, and $\kappa_{kj}=4\pi a_{kj}N/a_0$.
We can express as $\kappa_{kj}=a_0^2/2a_h^2$ using the healing length $a_h=\sqrt{8\pi a_{kj} N/a_0^3}$.  
Since the aspect ratio $\gamma_z$ of the trapping frequencies is typically small, it is more convenient for numerical studies to 
rescale the elongated $z$-axis as follows:
\begin{eqnarray}
 z^{\prime\prime}&=\gamma_z z^\prime,\\
 \psi^{\prime\prime}_i&=\frac{1}{\sqrt{\gamma_z}}\psi_i^\prime.
\end{eqnarray}
This rescaling finally yields our equations of motion:
\begin{eqnarray}
\fl   i\partial_t\psi_a&=\left(-\frac{1}{2}\left[\nabla^2_\perp + \gamma_z^2\partial_z^2 \right] + \frac{r^2}{2} + \gamma_z\kappa_{ja} |\psi_a|^2+\gamma_z\kappa_{jb}|\psi_b|^2\right)\psi_a
  +\frac{\Omega_a({\bf r})\Omega_b^*(z)}{8\omega\Delta}\psi_b\label{eq:rescaled_eqmo_final1}\\
\fl  i\partial_t\psi_b&=\left(-\frac{1}{2}\left[\nabla^2_\perp + \gamma_z^2\partial_z^2 \right] + \frac{r^2}{2} + \gamma_z\kappa_{ja} |\psi_a|^2+\gamma_z\kappa_{jb}|\psi_b|^2-\frac{\delta}{\omega}\right)\psi_b
  +\frac{\Omega_a^*({\bf r})\Omega_b(z)}{8\omega\Delta}\psi_a
  \label{eq:rescaled_eqmo_final2}
\end{eqnarray}
where we have dropped the primes for convenience. 
Furthermore, applying the same rescaling to the paraxial wave equation~[\refeq{eq:paraxial}], we find the following expression in 
our rescaled units:
\begin{equation}
2ik_0\partial_z \mathcal{E}({\bf r}_{\perp},z) = -\nabla^2_{\perp}\mathcal{E}({\bf r}_{\perp},z).
\label{eq:paraxial_rescaled}
\end{equation}
Here, $k_0$ has been redefined and stands for $k_0=2\pi a_0\gamma_z/\lambda$. 
Then, the functions defined in~\refeq{eq:LG_modes} remain solutions to~\refeq{eq:paraxial_rescaled} in the new coordinates, 
and using our modified $k_0$, we can still write the Rayleigh length as $z_r=k_0\sigma^2$, where 
$z_r$ is measured in units of $a_0/\gamma_z$ and $\sigma$ in units of $a_0$.

We have $m=1.44\times 10^{-25}$kg for ${}^{87}$Rb, so that a trapping frequency of $\omega=2\pi\times 10$Hz leads to 
an order of magnitude of $a_0=3.4\mu$m for the typical transverse extent of the BEC. 
An order of magnitude estimate for the off-diagonal coupling terms is given by 
\begin{equation}
 \frac{\Omega_a({\bf r})\Omega_b^*(z)}{8\omega\Delta}\approx \frac{\Omega_a({\bf r})\Omega_b^*(z)}{\Delta}0.016 {\mathrm s} \approx \Range{2d2}{2d6}
\end{equation}
for $\Omega_i\approx \Range{1}{10}$MHz and $\Omega_i/\Delta=\Range{0.01}{0.1}$. 
The required tight focusing of the beam leads to the fact, that we only need 
moderate beam powers to achieve adequate Rabi-frequencies. 
In the following, we set the two-photon detuning $\delta=0$ to achieve optimal transfer. 

Finally, we need to find an estimate for the Rabi pulse duration $t_d$. 
To this end, consider the simple case, where $\Omega_a$ and $\Omega_b$ describe driving by plane waves. 
We seek to find the optimal time for maximal transfer of atoms from $\ket{a}$ to $\ket{b}$. 
Given that the ``off-diagonal'' terms in the coupled equations~\refeqs{eq:rescaled_eqmo_final1}{eq:rescaled_eqmo_final2} are much larger than the diagonals, we can approximate 
the population $N_b$ in state $\ket{b}$ as 
\begin{equation}
N_b\sim 1-\cos^2\left(\frac{\Omega_a\Omega_b^*}{8\Delta \omega}t\right).
\end{equation}
This expression allows us to find the optimal (non-dimensional) time duration of the Rabi pulse $t_d$ that leads to a complete transfer of population:
\begin{equation}
 t_d=\frac{\pi/2}{\Omega_a\Omega_b^*/8\Delta \omega}.
 \label{eq:t_d}
\end{equation}
On the other hand, this consideration does not apply to our case due to the spatial dependence of the Rabi frequency $\Omega_a({\bf r})$. 
Due to spatial dependence of the intensity variations, we do not really have an optimal pulse duration 
$t_d$, and our choice of $t_d$ is a compromise between on one hand 
having sufficiently large transfer of atoms to state $\ket{b}$
and on the other hand avoiding population being transferred back from
state $\ket{b}$ to state $\ket{a}$ in positions where the intensity is large and thus the transfer 
dynamics is faster.
Hence, we have to choose shorter pulse durations than what~\refeq{eq:t_d} suggests. 
Furthermore, due to the nodal lines in the pump beam, it is never possible to use this arrangement for 
a total conversion of atomic population from $\ket{a}$ to $\ket{b}$. 
With these aspects kept in mind, we may still use~\refeq{eq:t_d} as a rough estimate for the order of magnitude for our choice of $t_d$.

\section{Numerical Investigation}

In the previous section, we elaborated on how to inscribe complex knotted vortex lines into matter waves. Whereas the 
simple argument of~\refeq{eq:simple_argument} allows us to conclude that our setup should work at least in the perturbative regime for 
short time-scales and small amounts of atoms in state $\ket{b}$, we cannot infer what happens beyond this perturbative regime. 
In this section, we aim to numerically study excitation and decay of our highly excited states into more elementary unknots or 
vanishing of the latter by collision and annihilation of vortex lines. 
We will do so by considering specific examples. However, these examples are by no means an exhaustive treatment of the
the huge potential manifold of realizations possible. To thoroughly understand the dynamics remains 
an elusive goal, which goes beyond the scope of this paper. 

\subsection{Hopf-link}

\begin{figure}\begin{center}
\includegraphics[width=\columnwidth]{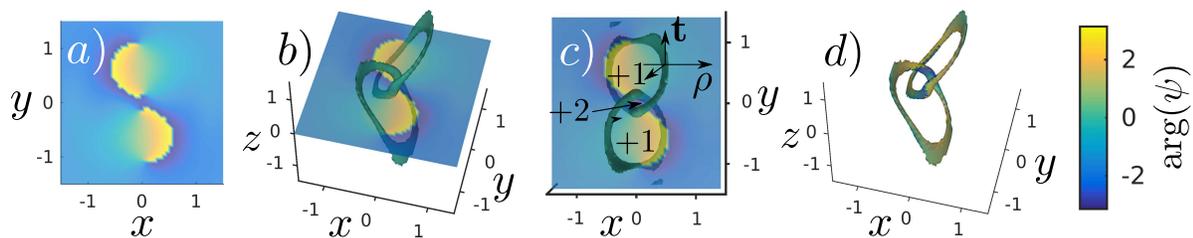}
\caption{
(color online) (a) Illustration of the phase of $\psi_b$ in the $(x,y)$-plane after short time evolution. (b) Additional to the phase in the $(x,y)$-plane from Fig. (a), 
the green surface represents a low-value isodensity surface of $|\psi_b|^2$ around the 
vortex line imprinted on the BEC. (c) Projection of Fig (b) into the $(x,y)$-plane and associated indices according to~\refeq{eq:I_j}. Expected dynamics should thus be, that 
the central region moves faster in the $z$-direction relative to its neighbouring regions. 
(d) shows again the isosurface from Fig. (b), however, the coloring of the isosurface illustrates the value of the phase at each position of the isosurface. 
\label{fig:hopf_projection}}
\end{center}\end{figure}

One of the simplest torus knots is the so-called Hopf-link, which corresponds to setting $n=2$ in~\refeq{eq:ansatz_light} and is 
depicted e.g. by the green isodensity surface in~\reffig{fig:hopf_projection}(b). Using the method outlined in~\ref{sec:knotted_light_field} and~\refeq{eq:determine_alp}, we find the following superposition 
coefficients of Laguerre-Gaussian polynomials for a Hopf-link:
\begin{eqnarray}
  \frac{\Omega_a\Omega_b^*}{8\omega\Delta} &= A[(1-2w^2+2w^4)\mathrm{LG}_{00}+(2w^2-4w^4)\mathrm{LG}_{01}+2w^4\mathrm{LG}_{02}\nonumber\\
  &-4\sqrt{2}w^2\mathrm{LG}_{20}]\Theta(t_{d}-t).
\end{eqnarray}
To this end, let us consider the dynamics for $\kappa_{ab}=4\pi a_{ab} N/a_0=1887$, $\gamma_z=0.125$. 
We set the two-photon detuning to $\delta=0$.
This can be achieved by using typical parameters of ${}^{87}$Rb with $\omega=2\pi\times 10 $Hz, $\omega_z=2.5\pi $Hz, $N=93000$, 
$a_{ab}=5.5$nm, $\lambda=780$nm, which amounts to typical 
units of length scales of $a_0=3.4\mu$m in the $(x,y)$-plane and $a_0/\gamma_z=27.2\mu m$ in the $z$-direction.
Thomas-Fermi like dynamics, where the nonlinearity is large compared to the broadening due to the Laplacian, 
is preferable to ensure robustness of the vortex cores and avoid the trivial broadening of the latter. 
For that reason, we have to choose a sufficiently large value for $\kappa_{ij}$.
For the other $a_{ij}$, we assume the ratio $a_{aa}$:$a_{ab}$:$a_{bb}$ to be given by $1.03$:$1$:$0.97$~\cite{Cornell:PRL:1998}.
With respect to our Rabi pulse, we choose 
$A=450$, $w=1.5$, $\sigma=0.675$ corresponding to $2.3\mu$m, 
and a pulse duration of $t_{d}=0.002$ corresponding to $31.8\mu$s.

We use the common Fourier split-step method~\cite{Agrawal_Kivshar:2003} and adaptive Runge-Kutta algorithm to fourth order for the time-step for numerical computation. 
Parts of the code were written using~\cite{XMDS}. To find the appropriate state $\psi_a(t=0)$, we used imaginary time-evolution (e.g.~\cite{Tosi:PRE:2000}).
It is crucial to use a variable time-step, since the dynamics during the time duration of the Rabi-pulse $t_d$ has to be temporally 
fully resolved, and thus the time-step has to be much smaller than $t_d$ within $t_d$. After $t>t_d$, the time-step can be chosen 
much larger, since it only needs to resolve the characteristic time-scale of the BEC dynamics. We used time-steps varying between $\Delta t=\Range{d-8}{5d-4}$ and
a spatial resolution of $\Delta x=0.05-0.06$ with $220^3$ points.

\reffig{fig:hopf_projection}(a-d) depict an isodensity surface of the condensate wave function $\psi_b$ with small value,  
which serves as an illustration of the knotted vortex core, and its phase profile in different ways, shortly after the imprinting occurred.
Here, the initial wave function $\psi_a(t=0)$ was computed as the ground state of the trap and $\psi_b(t=0)=0$. 

A basic qualitative anticipation of the dynamics can be found by 
projecting the knot orthogonally to the beam- or $z$-axis and considering the momentum associated with 
the (unit-)areas, as elaborated in~\cite{Ricca:Proc:2013}. Consider the projection in the $(x,y)$-plane, as 
shown in~\reffig{fig:hopf_projection}(c). 
The index $I_j$ assigned to each of the areas $R_j$ enclosed by the vortex lines $\gamma$ can be found by evaluating the sum~\cite{Ricca:Proc:2013}
\begin{equation}
 I_j=\sum_{{\bf \rho}\cap {\bf \gamma}} \mathrm{sign} \left( {\bf e_z} {\bm \rho} \times {\bf t} \right),
 \label{eq:I_j}
\end{equation}
which runs over all intersections of the chosen vector ${\bf\rho}$ with the projected vortex line $\gamma$ for a given region $R_j$. 
Here, ${\bm \rho}$ denotes an arbitrary vector pointing from the inside to the outside of the enclosed area 
and $\mathrm{sign}$ is the usual sign-function taking the values $\mathrm{sign}(\ast)=\pm 1$. 
Furthermore, the vector $\bf{t}$ is a tangent vector to the vortex line curve (denoted as $\gamma$) at the point where ${\bm \rho}$ crosses $\gamma$. 
The direction of $\bf t$ is given by the orientation of the arcs, which is determined by the phase~[see~\reffig{fig:hopf_projection}(c)].
The values assigned to the regions in~\reffig{fig:hopf_projection}(c) correspond to the values computed by~\refeq{eq:I_j}.
The associated momenta of a region $R_j$ can be found by~\cite{Ricca:Proc:2013}
\begin{equation}
 \left({\bf p}_z\right)_j=\oint_{R_j} {\bm \omega} d^2r I_j,
\end{equation}
where $\bm \omega$ denotes the vorticity. 

\begin{figure}\begin{center}
\includegraphics[width=\columnwidth]{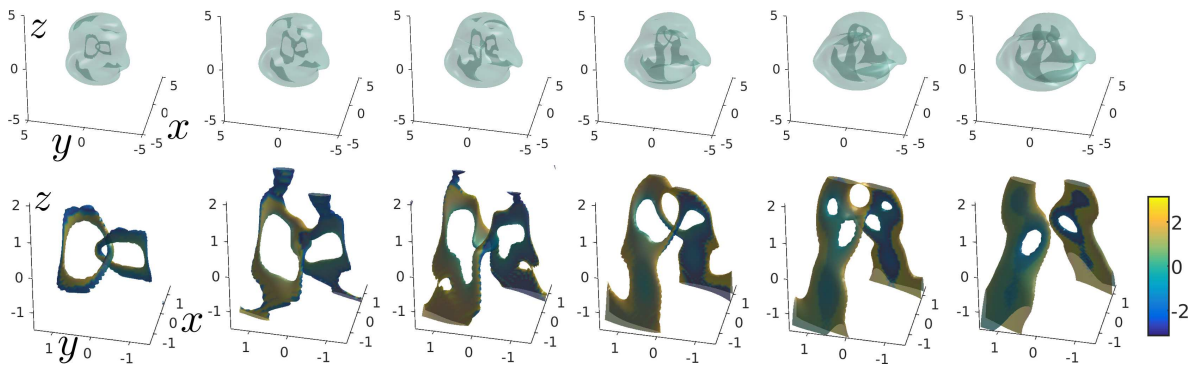}
\caption{
(color online) Dynamics of the condensate wavefunction $\psi_b$, whose vortex lines form a Hopf-link. Both upper and lower row show the dynamics in timesteps of 
$\Delta t=0.1$ (corresponding to $\Delta t=1.6$ms) starting from $t=0.05$ (corresponding to $t=0.8$ms). 
Both rows use the same isosurface for $|\psi_b|^2$, however, in the lower row we cropped the box at a smaller value, so that the ``boundary'' of
the condensate does not obfuscate the dynamics of the vortex lines. Additional to that, we used the phase to color the isosurface in the lower row. 
We can see an overall drift the positive $z$-direction, which can be understood 
from~\reffig{fig:hopf_projection}. 
See also the movies \href{hopf_1.mp4}{\textit{hopf\_1.mp4}} and \href{hopf_2.mp4}{\textit{hopf\_2.mp4}} for the full propagation dynamics.
\label{fig:hopf_to_unknot}}
\end{center}\end{figure}

What one can deduce from these arguments is a movement in the positive $z$-direction with the central region traveling fastest [\reffig{fig:hopf_projection}(c)]. 
Let us now look at the numerically computed dynamics. 
Snapshots of the latter are shown in~\reffig{fig:hopf_to_unknot} (see also movies \href{hopf_1.mp4}{\textit{hopf\_1.mp4}} and \href{hopf_2.mp4}{\textit{hopf\_2.mp4}}).
Clearly, there is an overall center-of-mass mass motion towards the positive $z$-direction, as expected. 
However, reconnection of vortex lines as well as finite size effects of the condensate leading to sharp gradients in the density lead to a very involved dynamics, 
which we were not able to predict or understand in simple terms.
Surprisingly, the formation decays into two unknots that propagate into the positive and negative $y$-direction, respectively. 

\subsection{Trefoil}\label{sec:trefoil}

In this section we will study the dynamics of a trefoil imprinted on a BEC. The dynamics is generic, quite different initial conditions lead to similar results,
the basics of which can be understood in the simple terms described before. 

In this case, let $w=1.2$. Instead of using the prefactors $a_{l,p}(w)$ from~\refeq{eq:determine_alp}, 
we use the optimized prefactors given in~\cite{King:thesis,Dennis:Nature:2010} of the Laguerre-Gaussian modes for the light field.  
\begin{equation}
 \fl \frac{\Omega_a\Omega_b^*}{8\omega\Delta} = A( 1.51\mathrm{LG}_{00} - 5.06\mathrm{LG}_{10} + 7.23\mathrm{LG}_{20}- 2.03\mathrm{LG}_{30} - 3.97\mathrm{LG}_{03} )\Theta(t_{d}-t).
\end{equation}
To this end, let us use $\sigma=0.78$ corresponding to $2.65\mu$m and $A=700$. 
Furthermore, let us choose $\kappa = 3753$, $\gamma_z = 0.05$. 
This can be realized using again $\omega=2\pi\times 10 $Hz, $\omega_z=1\pi $Hz,
$N=1.85\times10^5$, $a_{ab}=5.5$nm and $\lambda=780$nm. 
With respect to our Rabi pulse, we use
a pulse duration of $t_{d}=0.004$ corresponding to $63.7\mu$s.

\begin{figure}\begin{center}
\includegraphics[width=\columnwidth]{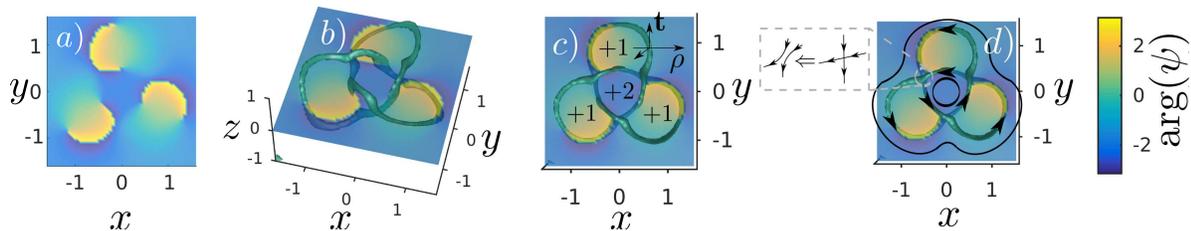}
\caption{
(color online) (a) Illustration of the phase of $\psi_b$ in the $(x,y)$-plane after short time evolution. (b) Additional to the phase in the $(x,y)$-plane from Fig. (a), 
the green line represents a low-value isodensity surface of $|\psi_b|^2$ around the 
vortex line imprinted on the BEC. (c) Projection of Fig (b) into the $(x,y)$-plane and associated indices. 
Since the indices reflect the momenta of the regions, we expect that 
the central region moves faster in the $z$-direction relative to its neighboring regions. 
(d) shows the expected breakup of the green vortex lines 
into the solid black lines. The black arrows illustrate the orientation, which is determined by the phase. The grey dashed inset illustrates 
the expected product of collision and reconnection of the vortex lines according to, e.g.~\cite{Koplik:PRL:1993}.
\label{fig:trefoil_projection}}
\end{center}\end{figure}

\reffig{fig:trefoil_projection}(a) illustrates the phase of $\psi_b$ and (b) its vortex core shortly after imprinting. 
Projecting onto the $(x,y)$-plane and assigning values according to~\refeq{eq:I_j} leads to~\reffig{fig:trefoil_projection}(c).
Again, we can expect that the overall knot will propagate in the positive $z$-direction, and the central part carries the largest momentum.
\reffig{fig:trefoil_projection}(d) illustrates the expected reconnection dynamics according to the rules found in e.g.~\cite{Koplik:PRL:1993}.
The arrows indicate the direction of ${\bf t}$, which is again determined by the phase. The inset (grey dashed box) illustrates how 
the vortex lines should reconnect. If we apply this simple rule to all three crossings, we can expect that the green vortex line evolves into what is depicted by the solid 
black line, i.e. a decay into two unkots.

\begin{figure}\begin{center}
\includegraphics[width=\textwidth]{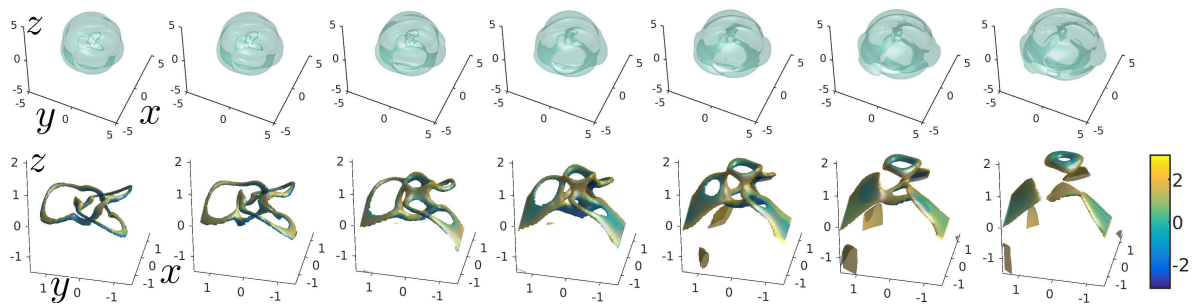}
\caption{
(color online) Dynamics of the condensate wavefunction $\psi_b$, whose vortex lines form a trefoil knot. Both upper and lower row show the same dynamics in timesteps of 
$\Delta t=0.1$ (corresponding to $\Delta t=0.8$ms) starting from $t=0.05$ (corresponding to $t=0.8$ms). 
Both rows use the same isosurface for $|\psi_b|^2$, however, in the lower row we cropped the box at a smaller value, so that the ``boundary'' of
the condensate does not obfuscate the dynamics of the vortex lines. Additional to that, we used the phase to color the isosurface in the lower row. 
Upon evolution, vortex cores reconnect, which leads to a decay into a vortex ring being expelled from 
the central region, which can be understood from~\reffig{fig:trefoil_projection}).
See also the movie \href{trefoil_1.mp4}{\textit{trefoil\_1.mp4}} and \href{trefoil_2.mp4}{\textit{trefoil\_2.mp4}} for the full propagation dynamics.
\label{fig:vortex_expelled}}
\end{center}\end{figure}

Let us now consider the dynamics, which is shown in~\reffig{fig:vortex_expelled} (see also the movie 
\href{trefoil_1.mp4}{\textit{trefoil\_1.mp4}} and \href{trefoil_2.mp4}{\textit{trefoil\_2.mp4}} for full dynamics).
We see that the vortex lines of this specific trefoil knot first reconnect [see~\reffig{fig:vortex_expelled}], and the 
central regions travels fastest into the positive $z$-direction according to our expectation~\reffig{fig:trefoil_projection}(b--c).
After that, the central region expels an unknot or vortex ring which decouples from the rest of the knot, as shown in~\reffig{fig:vortex_expelled}. 
The single unknot propagates to the positive $z$-direction faster than the remaining part of the knot. 
Interestingly, our dynamics differs from the usual dynamics 
of a clear breakup into two unknots as expected from [\reffig{fig:trefoil_projection}(c), solid black line] 
and what has been found in~\cite{Barenghi:PRE:2012,Irvine:arXiv:2015}. 
Differences to previous observations are due to finite size effects of the cloud, that in our case 
the knot has a large aspect ratio, which with our rescaling basically means that the dynamics in the $z$-direction is much slower, smallness of the knot, and finally that we 
actually consider two coupled fields (dynamics of $\psi_a$ not shown). 

\section{Conclusions}
Recently there has been a growing interest in the dynamics of knotted vortex lines in BECs. 
However, thus far there has been no actual proposal on how to excite such matter waves in a controlled fashion. 
In this work, we have presented a setup with physically realistic parameters to create such knotted vortex lines in ultracold matter waves. 
We combined recent theoretical and experimental results from complex light shaping which allowed us to create knotted vortex lines embedded in light fields. We 
use the latter as a probe field for three-level atoms in a $\Lambda$-type setup to inscribe the nodal lines to BECs. 
The setup is quite generic, and allows investigation of a large variety of potential dynamics. 
The finite time span to populate state $\psi_b$ reduces production of soundwaves. 
The finiteness of the condensate, the smallness of the knot as well as the reconnections of the vortex lines give
rise to a very involved dynamics. 
The velocity in the $z$-direction of the vortex knot should not be large.  
This velocity can be controlled by the ratio between the trapping frequencies $\gamma_z=\omega_z/\omega$ relative to the condensate.  
The probe field which determines $\Omega_a$ should not be too spatially broad, otherwise the value of $\gamma_z$ required to fully embed the knotted vortex line 
becomes impractical. We have assumed that the light field can be treated within the paraxial approximation, although we note that, in principle, 
this condition can  be relaxed to non-paraxial knotted light fields~\cite{Dennis:OL:11}. 
Due to the tight focusing of the optical beam, only moderate powers of the light fields that determine the Rabi-frequencies $\Omega_i$ are required. 
Thus, the setup should work generically for a large range of parameters. 
We have shown two illustrative examples of a trefoil knot and a Hopf-link, and discussed the dynamics using already well-established techniques. 
The data presented in this paper is available online at~\cite{Maucher:data}.

\section{Acknowledgements} 
This work was funded by the Leverhulme Trust Research Programme Grant RP2013-K-009, SPOCK: Scientific Properties Of Complex Knots. 
We would like to thank P. M. Sutcliffe, J. L. Helm, T. P. Billam, D. Sugic and M. Dennis for stimulating discussions. 

\section*{References}
\bibliographystyle{unsrt}
\bibliography{bib}

\begin{thebibliography}{10}

\bibitem{Berry:royal:1974}
J.~F. Nye and M.~V. Berry.
\newblock Dislocations in wave trains.
\newblock {\em Proc. R. Soc. London Ser. A-Math. Phys. Eng. Sci.}, 336:165,
  1974.

\bibitem{Irvine:NatPhys:2008}
W.~T.~M. Irvine and D.~Bouwmeester.
\newblock Linked and knotted beams of light.
\newblock {\em Nature Phys.}, 4:716–720, 2008.

\bibitem{knot_book}
C.~Adams.
\newblock {\em The Knot Book}.
\newblock American Mathematical Soc., 1994.

\bibitem{Shabtay:OC:2003}
G.~Shabtay.
\newblock Three-dimensional beam forming and {E}wald’s surfaces.
\newblock {\em Opt. Commun.}, 226:33, 2003.

\bibitem{Padgett:NJP:2005}
J.~Leach, M.~R. Dennis, J.~Courtial, and M.~J. Padgett.
\newblock Vortex knots in light.
\newblock {\em New J. Phys.}, 7:55, 2005.

\bibitem{Whyte:NJP:2005}
G.~Whyte and J.~Courtial.
\newblock Experimental demonstration of holographic three-dimensional light
  shaping using a {G}erchberg-{S}axton algorithm.
\newblock {\em New J. Phys.}, 7:117, 2005.

\bibitem{Dennis:Nature:2010}
M.~R. Dennis, R.~P. King, Barry Jack, K~O'Holleran, and M.~J. Padgett.
\newblock Isolated optical vortex knots.
\newblock {\em Nat. Phys.}, 6:118, 2010.

\bibitem{Shanblatt:OE:11}
E.~R. Shanblatt and D.~G. Grier.
\newblock Extended and knotted optical traps in three dimensions.
\newblock {\em Opt. Express}, 19:5833, 2011.

\bibitem{Moffatt:JFM:1969}
H.~K. Moffatt.
\newblock The degree of knottedness of tangled vortex lines.
\newblock {\em J. of Fluid Mech.}, 35:117--129, 1 1969.

\bibitem{Moffatt:nature:1990}
H.~K. Moffatt.
\newblock The energy spectrum of knots and links.
\newblock {\em Nature (London)}, 347:367, 1990.

\bibitem{Paul:PRE:2003}
P.~M. Sutcliffe and A.~T. Winfree.
\newblock Stability of knots in excitable media.
\newblock {\em Phys. Rev. E}, 68:016218, 2003.

\bibitem{Sutcliffe:PRL:2016}
F.~Maucher and P.~Sutcliffe.
\newblock Untangling knots via reaction-diffusion dynamics of vortex strings.
\newblock {\em Phys. Rev. Lett.}, 116:178101, 2016.

\bibitem{Tkalec:science:2011}
U.~Tkalec, M.~Ravnik, S.~Copar, S.~Zumer, and I.~Musevic.
\newblock Reconfigurable knots and links in chiral nematic colloids.
\newblock {\em Science}, 333:62--65, 2011.

\bibitem{Babaev:PRL:2002}
E.~Babaev.
\newblock Dual neutral variables and knot solitons in triplet superconductors.
\newblock {\em Phys. Rev. Lett.}, 88:177002, 2002.

\bibitem{Babaev:PRB:2009}
E.~Babaev.
\newblock Non-{M}eissner electrodynamics and knotted solitons in two-component
  superconductors.
\newblock {\em Phys. Rev. B}, 79:104506, 2009.

\bibitem{Kelvin:1867}
W.~Thompson (Lord~Kelvin).
\newblock On vortex atoms.
\newblock {\em P. Roy. Soc. Edinb. A}, 6:94--105, 1869.

\bibitem{Faddeev:Nature:1997}
L.~Faddeev and A.~J. Niemi.
\newblock Stable knot-like structures in classical field theory.
\newblock {\em Nature}, 387:58–61, 1997.

\bibitem{Paul:PRL:1998}
R.~A. Battye and P.~M. Sutcliffe.
\newblock Knots as stable soliton solutions in a three-dimensional classical
  field theory.
\newblock {\em Phys. Rev. Lett.}, 81:4798--4801, 1998.

\bibitem{TinLun:PRL:98}
Tin-Lun Ho.
\newblock Spinor {B}ose condensates in optical traps.
\newblock {\em Phys. Rev. Lett.}, 81:742--745, 1998.

\bibitem{Machida:JphysJpn:1998}
T.~Ohmi and K.~Machida.
\newblock Bose-einstein condensation with internal degrees of freedom in alkali
  atom gases.
\newblock {\em J. Phys. Soc. Jpn.}, 67:1822, 1998.

\bibitem{Niemi:PRB:2002}
E.~Babaev, L.~D. Faddeev, and A.~J. Niemi.
\newblock Hidden symmetry and knot solitons in a charged two-condensate {B}ose
  system.
\newblock {\em Phys. Rev. B}, 65:100512, 2002.

\bibitem{Ueda:PRL:08}
Y.~Kawaguchi, M.~Nitta, and M.~Ueda.
\newblock Knots in a spinor {B}ose-{E}instein condensate.
\newblock {\em Phys. Rev. Lett.}, 100:180403, 2008.

\bibitem{Ueda:PTPS:2010}
Y.~Kawaguchi, M.~Nitta, and M.~Ueda.
\newblock Topological excitations in spinor {B}ose-{E}instein condensates.
\newblock {\em Prog. Theor. Phys. Suppl.}, 186:455, 2010.

\bibitem{Mottonen:NatPhys:2016}
D. S. Hall, M. W. Ray, K.~Tiurev, E.~Ruokokoski, A. H. Gheorghe, and
  M.~M{\"o}tt{\"o}nen.
\newblock Tying quantum knots.
\newblock {\em Nat. Phys.}, 2016.

\bibitem{Speight:JGP:2010}
J.~M. Speight.
\newblock Supercurrent coupling in the {F}addeev-{S}kyrme model.
\newblock {\em J. Geom. Phys.}, 60:599, 2010.

\bibitem{Barenghi:PRE:2012}
D.~Proment, M.~Onorato, and C.~F. Barenghi.
\newblock Vortex knots in a {B}ose-{E}instein condensate.
\newblock {\em Phys. Rev. E}, 85:036306, 2012.

\bibitem{Barenghi:JOP:2014}
D.~Proment, M.~Onorato, and C.~F. Barenghi.
\newblock Torus quantum vortex knots in the {G}ross-{P}itaevskii model for
  {B}ose-{E}instein condensates.
\newblock {\em Journal of Physics: Conference Series}, 544:012022, 2014.

\bibitem{Frisch:PRL:1992}
T.~Frisch, Y.~Pomeau, and S.~Rica.
\newblock Transition to dissipation in a model of superflow.
\newblock {\em Phys. Rev. Lett.}, 69:1644--1647, 1992.

\bibitem{Koplik:PRL:1993}
J.~Koplik and H.~Levine.
\newblock Vortex reconnection in superfluid helium.
\newblock {\em Phys. Rev. Lett.}, 71:1375--1378, 1993.

\bibitem{Bewley:PNAS:2008}
G.~P. Bewley, M.~S. Paoletti, K.~R. Sreenivasan, and D.~P. Lathrop.
\newblock Characterization of reconnecting vortices in superfluid helium.
\newblock {\em Proceedings of the National Academy of Sciences}, 105:13707,
  2008.

\bibitem{Barenghi:PhysFluid:2012}
S.~Zuccher, M.~Caliari, A.~W. Baggaley, and C.~F. Barenghi.
\newblock Quantum vortex reconnections.
\newblock {\em Physics of Fluids}, 24:125108, 2012.

\bibitem{Barenghi:PRL:2008}
S.~Z. Alamri, A.~J. Youd, and C.~F. Barenghi.
\newblock Reconnection of superfluid vortex bundles.
\newblock {\em Phys. Rev. Lett.}, 101:215302, 2008.

\bibitem{Leadbeater:PRL:2001}
M.~Leadbeater, T.~Winiecki, D.~C. Samuels, C.~F. Barenghi, and C.~S. Adams.
\newblock Sound emission due to superfluid vortex reconnections.
\newblock {\em Phys. Rev. Lett.}, 86:1410--1413, 2001.

\bibitem{Ruostekoski:PRA:2005}
J.~Ruostekoski and Z.~Dutton.
\newblock Engineering vortex rings and systems for controlled studies of vortex
  interactions in {B}ose-{E}instein condensates.
\newblock {\em Phys. Rev. A}, 72:063626, 2005.

\bibitem{Wright:PRL:1997}
K.-P. Marzlin, W.~Zhang, and E.~M. Wright.
\newblock Vortex coupler for atomic {B}ose-{E}instein condensates.
\newblock {\em Phys. Rev. Lett.}, 79:4728--4731, 1997.

\bibitem{Dum:prl:98}
R.~Dum, J.~I. Cirac, M.~Lewenstein, and P.~Zoller.
\newblock Creation of dark solitons and vortices in {B}ose-{E}instein
  condensates.
\newblock {\em Phys. Rev. Lett.}, 80:2972--2975, 1998.

\bibitem{Ruostekoski:PRL:2004}
Z.~Dutton and J.~Ruostekoski.
\newblock Transfer and storage of vortex states in light and matter waves.
\newblock {\em Phys. Rev. Lett.}, 93:193602, 2004.

\bibitem{Andersen:PRL:2006}
M.~F. Andersen, C.~Ryu, Pierre Clad\'e, Vasant Natarajan, A.~Vaziri,
  K.~Helmerson, and W.~D. Phillips.
\newblock Quantized rotation of atoms from photons with orbital angular
  momentum.
\newblock {\em Phys. Rev. Lett.}, 97:170406, 2006.

\bibitem{Cornell:PRL::1999_2}
M.~R. Matthews, B.~P. Anderson, P.~C. Haljan, D.~S. Hall, M.~J. Holland, J.~E.
  Williams, C.~E. Wieman, and E.~A. Cornell.
\newblock Watching a superfluid untwist itself: Recurrence of {R}abi
  oscillations in a {B}ose-{E}instein condensate.
\newblock {\em Phys. Rev. Lett.}, 83:3358--3361, 1999.

\bibitem{Cornell:PRL::1999}
M.~R. Matthews, B.~P. Anderson, P.~C. Haljan, D.~S. Hall, C.~E. Wieman, and
  E.~A. Cornell.
\newblock Vortices in a {B}ose-{E}instein condensate.
\newblock {\em Phys. Rev. Lett.}, 83:2498--2501, 1999.

\bibitem{Schmiegelow:ARXIV:2015}
C.~T. Schmiegelow, J.~Schulz, H.~Kaufmann, T.~Ruster, U.~G. Poschinger, and
  F.~Schmidt-Kaler.
\newblock Excitation of an atomic transition with a vortex laser beam.
\newblock {\em arXiv:1511.07206}, 2015.

\bibitem{Cornell:PRL:2000}
B.~P. Anderson, P.~C. Haljan, C.~E. Wieman, and E.~A. Cornell.
\newblock Vortex precession in {B}ose-{E}instein condensates: Observations with
  filled and empty cores.
\newblock {\em Phys. Rev. Lett.}, 85:2857--2860, 2000.

\bibitem{Irvine:arXiv:2015}
D.~Kleckner, L.~H. Kauffman, and W.~T.~M. Irvine.
\newblock How superfluid vortex knots untie.
\newblock {\em Nat. Phys.}, 2016.

\bibitem{Cornell:PRL:1998}
D.~S. Hall, M.~R. Matthews, J.~R. Ensher, C.~E. Wieman, and E.~A. Cornell.
\newblock Dynamics of component separation in a binary mixture of
  {B}ose-{E}instein condensates.
\newblock {\em Phys. Rev. Lett.}, 81:1539--1542, 1998.

\bibitem{Milnor_book}
J.~Milnor.
\newblock {\em Singular Points of Complex Hypersurfaces}.
\newblock Princeton Univ. Press., 1969.

\bibitem{King:thesis}
R.~P. King.
\newblock {\em Knotting of Optical Vortices}.
\newblock PhD thesis, University of Southampton, 2010.

\bibitem{Alexander:PNAS:1923}
J.~W. Alexander.
\newblock A lemma on systems of knotted curves.
\newblock {\em P. Natl. Acad. Sci. USA}, 3:93–95, 1923.

\bibitem{Agrawal_Kivshar:2003}
Y.~S. Kivshar and G.~Agrawal.
\newblock {\em Optical Solitons: From Fibers to Photonic Crystals}.
\newblock Academic Press, San Diego, 2003.

\bibitem{XMDS}
G.~R. Dennis, J.~J. Hope, and M.~T. Johnsson.
\newblock Xmds2: Fast, scalable simulation of coupled stochastic partial
  differential equations.
\newblock {\em Computer Physics Communications}, 184:201, 2013.

\bibitem{Tosi:PRE:2000}
M.~L. Chiofalo, S.~Succi, and M.~P. Tosi.
\newblock Ground state of trapped interacting {B}ose-{E}instein condensates by
  an explicit imaginary-time algorithm.
\newblock {\em Phys. Rev. E}, 62:7438--7444, 2000.

\bibitem{Ricca:Proc:2013}
R.~L. Ricca.
\newblock Impulse of vortex knots from diagram projections.
\newblock {\em Procedia IUTAM}, 7:21--28, 2013.

\bibitem{Dennis:OL:11}
M.~R. Dennis, J.~B. G\"{o}tte, R.~P. King, M.~A. Morgan, and M.~A. Alonso.
\newblock Paraxial and nonparaxial polynomial beams and the analytic approach
  to propagation.
\newblock {\em Opt. Lett.}, 36(22):4452--4454, 2011.

\bibitem{Maucher:data}
F.~Maucher, S.~A. Gardiner, and I.~G. Hughes.
\newblock \url{http://collections.durham.ac.uk/files/1n79h428d}, and
  \url{http://dx.doi.org/10.15128/1n79h428d}, 2015.
\newblock {``Excitation of knotted vortex lines in matter waves: Supporting
  data''}.

\end{thebibliography}

\end{document}